# Systematic Study of LED Stimulated Recovery of Radiation Damage in Optical Materials

**K. K. Sahbaz[a,b,c] , B. Bilki[a,c,d,1], H. Dapo[c], I. G. Karslioglu[a,b,c], C. Kaya[c], M. Kaya[c], and M. Tosun[a,b,c]**

[a] *Beykent University, Istanbul, Turkey*
[b] *Ankara University, Ankara, Turkey*
[c] *Turkish Accelerator and Radiation Laboratory, Ankara, Turkey*
[d] *University of Iowa, Iowa City, USA*

*E-mail*: Burak.Bilki@cern.ch

ABSTRACT: The radiation damage in optical materials mostly manifests itself as the loss of optical transmittance. The optical materials recover from radiation damage to some extent when the radiation exposure is stopped. The recovery is at a faster rate in the presence of stimulating light. On the other hand, a systematic study of the dynamics of the recovery as a function of the stimulating light parameters such as its wavelength, intensity and exposure duration and method has not been performed in detail yet.

We established an LED recovery station which provides pulsed and continuous light at various wavelengths at custom geometries. We irradiated soda lime glass samples at a rate of 87.5 Gy/min to a total dose of 3.5 kGy and 7.0 kGy. The optical transmittance of the samples were then measured in 200 nm – 1500 nm range for an extended period of time. The recovery from radiation damage is improved, both in terms of timing and quantity, as the wavelength of the stimulating light decreases. Around 50 % improvement was measured both in recovery rate and the permanent damage when UV LED with a wavelength of 396 nm was used for stimulation. The trend is such that wavelengths deeper in the UV range would result in faster and more effective recovery from radiation damage. The LED stimulated recovery technique from radiation damage is a feasible implementation for the optical active media of radiation and particle detectors which operate in high radiation environments.

---

[1] Corresponding author

## Contents

## 1. Introduction

Optical materials, including scintillators, glasses and crystals, are widely used as active media of radiation detectors, as optical windows and as transparent insulating materials in collider detectors, beamlines and various parts of scientific facilities. The particular concern related to these optical media is the loss of optical transmittance with increased doses of radiation exposure. Recent findings on the effect of the dose rate, in addition to the total dose, on the radiation damage to the optical materials indicate that the dose rate should also be considered when designing optical detector systems for future implementations, further complicating the design of the detector systems [1, 2].

The loss of optical transmittance results in significant degradation of the performance of the optical detectors. For some materials, this loss can recover by itself within a certain amount of time and to a limited extent [3]. Some external stimuli, such as sunlight and LED light exposure, accelerate this process and enhance the maximum recovery rate. Irradiated samples stimulated with RGB LEDs were observed to have an accelerated recovery process and a higher final optical transmittance compared to the samples kept in natural conditions [4]. On the other hand, a systematic study of the dynamics of the recovery as a function of the stimulating light parameters has not been performed in detail so far. In order to perform this systematic study, we established an LED recovery station which provides light at various wavelengths and compared its performance across a dark box and an ambient light recovery condition.

Here we report on the details of the irradiation and recovery setups, and the results of recovery of optical materials from radiation damage under different mechanisms of stimulation.

## 2. Experimental Setup

In order to initiate the systematic studies of recovery mechanisms from radiation damage in optical materials, we irradiated two sets of three pieces of soda-lime float glass samples at the Medical Linac Facility of Turkish Accelerator and Radiation Laboratory (TARLA) [5, 6]. Figure 1 shows a sketch of the TARLA facility.

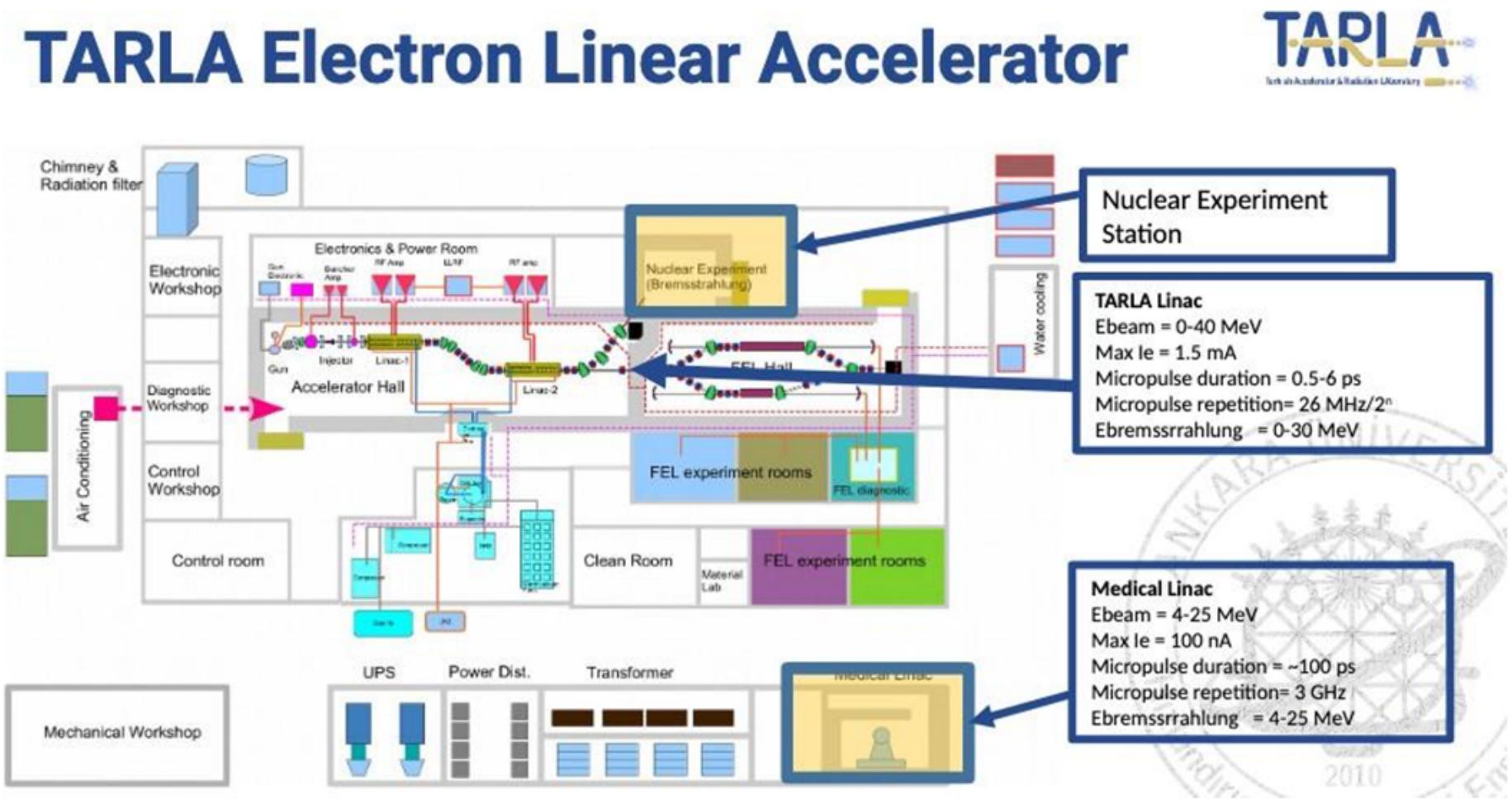

**Figure 1.** A sketch of the TARLA facility.

The Medical Linac provides 6-21 MeV bremsstrahlung photons. During acceleration, the energy spread of the electrons is about 20 %, and 1 % after the selection magnets. The conversion target is a 0.3 mm tungsten foil (90 % tungsten + 10 % rhenium). Several shaping filters downstream the target are used to create a spatially uniform photon beam. The beam is approximately 10 cm diameter at the experimental location and the lateral uniformity is within 3 %. The machine is designed to maintain a constant preset dose rate, which can be adjusted up to a maximum of 87.5 Gy/min.; therefore, as the environmental and machine conditions change, the electron gun current is automatically adjusted. In this study, 6 MeV bremsstrahlung photons were used at a dose rate of 87.5 Gy/min.

Two sets of three pieces of soda-lime float glass samples were irradiated. The total absorbed dose of the two sets of samples were 3.5 kGy and 7.0 kGy. Immediately following the irradiation, one sample from each set was placed: in a dark box; in a room with ambient light; at the LED recovery station. The samples had different thicknesses, which are shown in Table 1. Since the effect of the total dose on the loss of transmittance is relatively well known, the 3.5 kGy irradiated sample that was placed in the LED station was chosen to be a thicker sample in order to probe the effect of the sample thickness and the total absorbed dose on the recovery at the same time. The difference in the thicknesses of the 2.8 mm glasses is considered in the calculation of the systematic error rather than applying individual corrections. The 3.7 mm thick glass sample recovery is studied independently.

**Table 1.** The thicknesses of the soda-lime float glass samples used.

| Sample | Thickness (mm) |
|---|---|
| Dark Box - 3.5 kGy | 2.828 ± 0.014 |
| Ambient Light - 3.5 kGy | 2.773 ± 0.011 |
| LED Station - 3.5 kGy | 3.702 ± 0.016 |
| Dark Box – 7.0 kGy | 2.789 ± 0.011 |
| Ambient Light – 7.0 kGy | 2.749 ± 0.019 |
| LED Station – 7.0 kGy | 2.810 ± 0.017 |

The LED recovery station provides pulsed and continuous light at various wavelengths at custom locations. The station contains 5 different LEDs: ultraviolet (UV), blue, white, green and red. Figure 2 shows the spectra of the LEDs used in one of the recovery stations. The intensities of all the LEDs were brought closer together by designing custom electronics in order to minimize the systematics related to the operational parameters of the LEDs. The maximum variation between the intensities of the LEDs was 7 %. The results were compatible for the other LED recovery station. The LED stations were operated in pulsed mode at 10 Hz throughout this study.

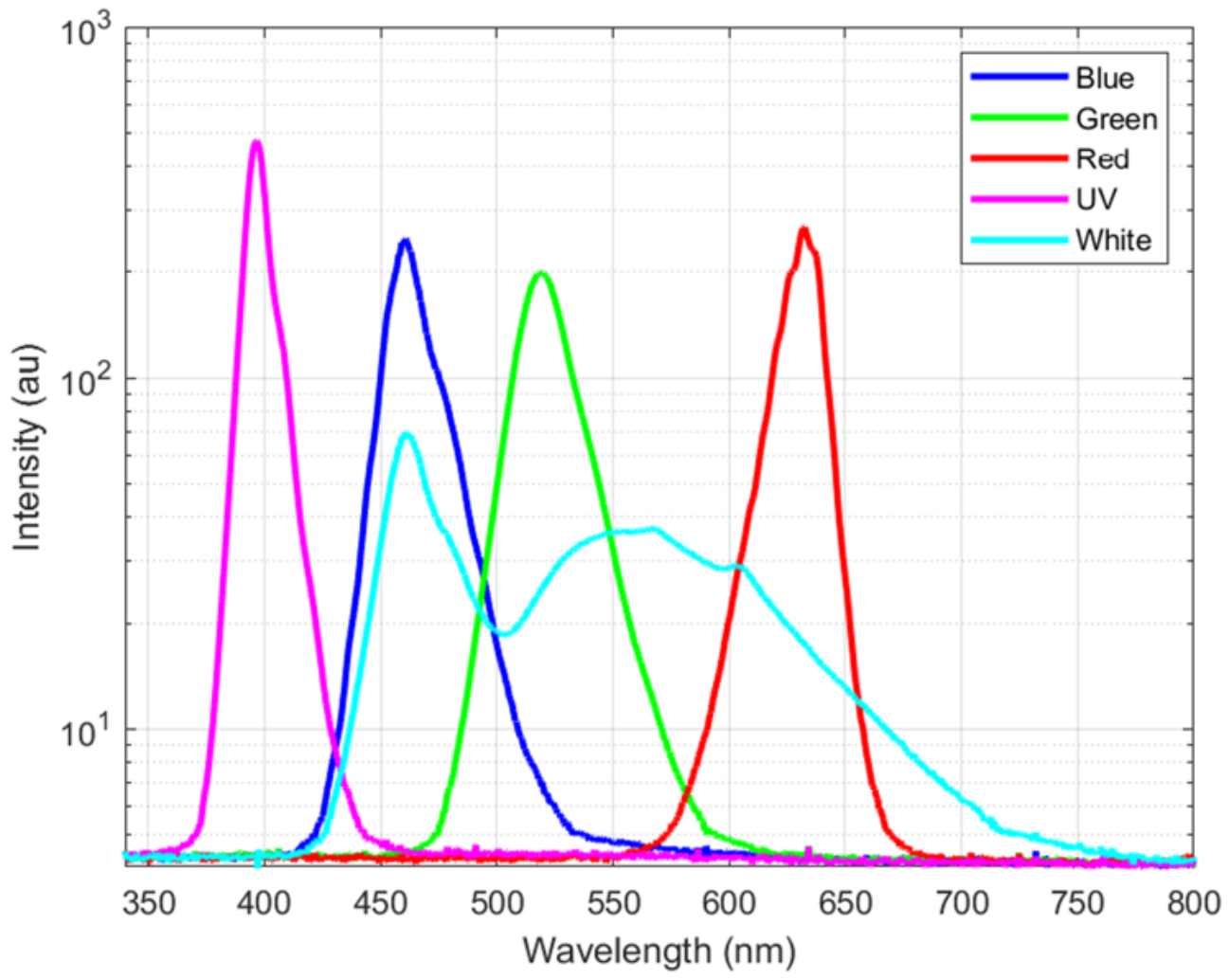

**Figure 2.** The spectra of the LEDs used in the recovery station.

Figure 3 shows pictures of the LED station (left) and an irradiated glass sample with the recovery and measurement template attached on top (right). The template defines five distinct and optically isolated locations on the glass samples which are then illuminated with individual LEDs. In order to minimize the reflection on the back side of the glass, the assembly was placed upside down on a black, non-reflective paper. Visual examination of the samples long after the irradiation shows sharp boundaries of the illumination windows.

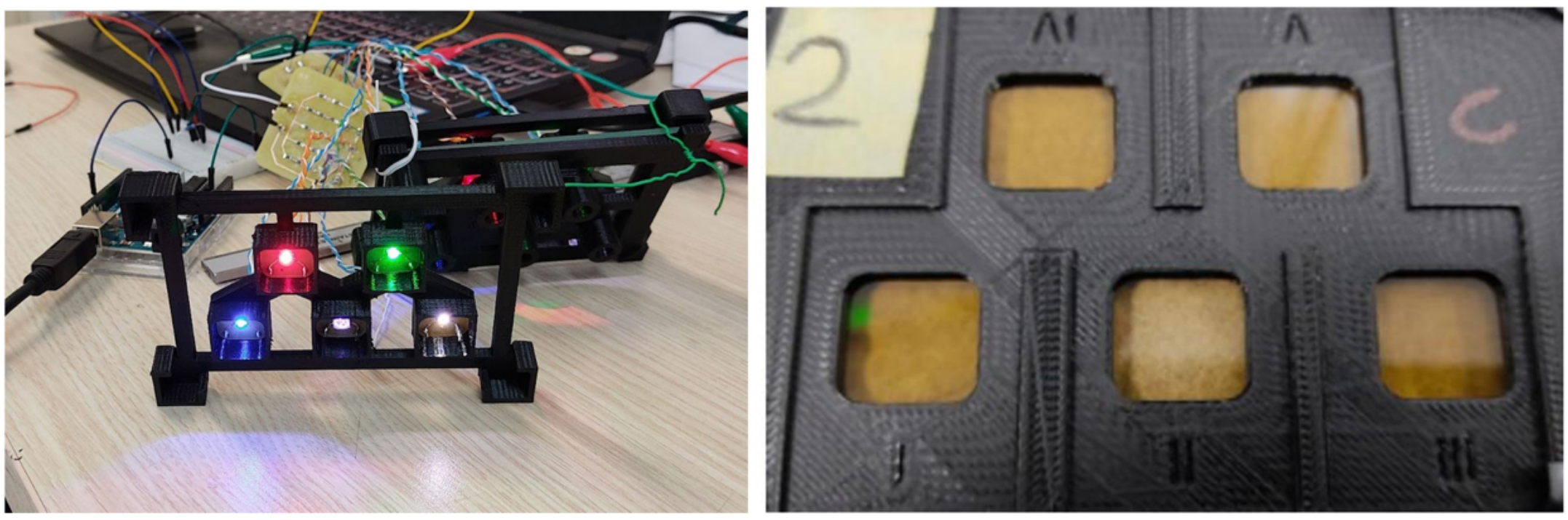

**Figure 3.** Pictures of the LED station (left) and an irradiated glass sample with the recovery and measurement template attached on top (right).

The LED recovery stations were operated in a temperature controlled room in order to provide a stable light intensity on the samples. Figure 4 shows the variation of the integral of the intensity for the UV LED and the room temperature for two consecutive days. The day-night temperature variation was approximately 2 °C and the fluctuations in the integral intensity was within 3 %. The result of the measurement with the UV LED was considered as common for all the LEDs.

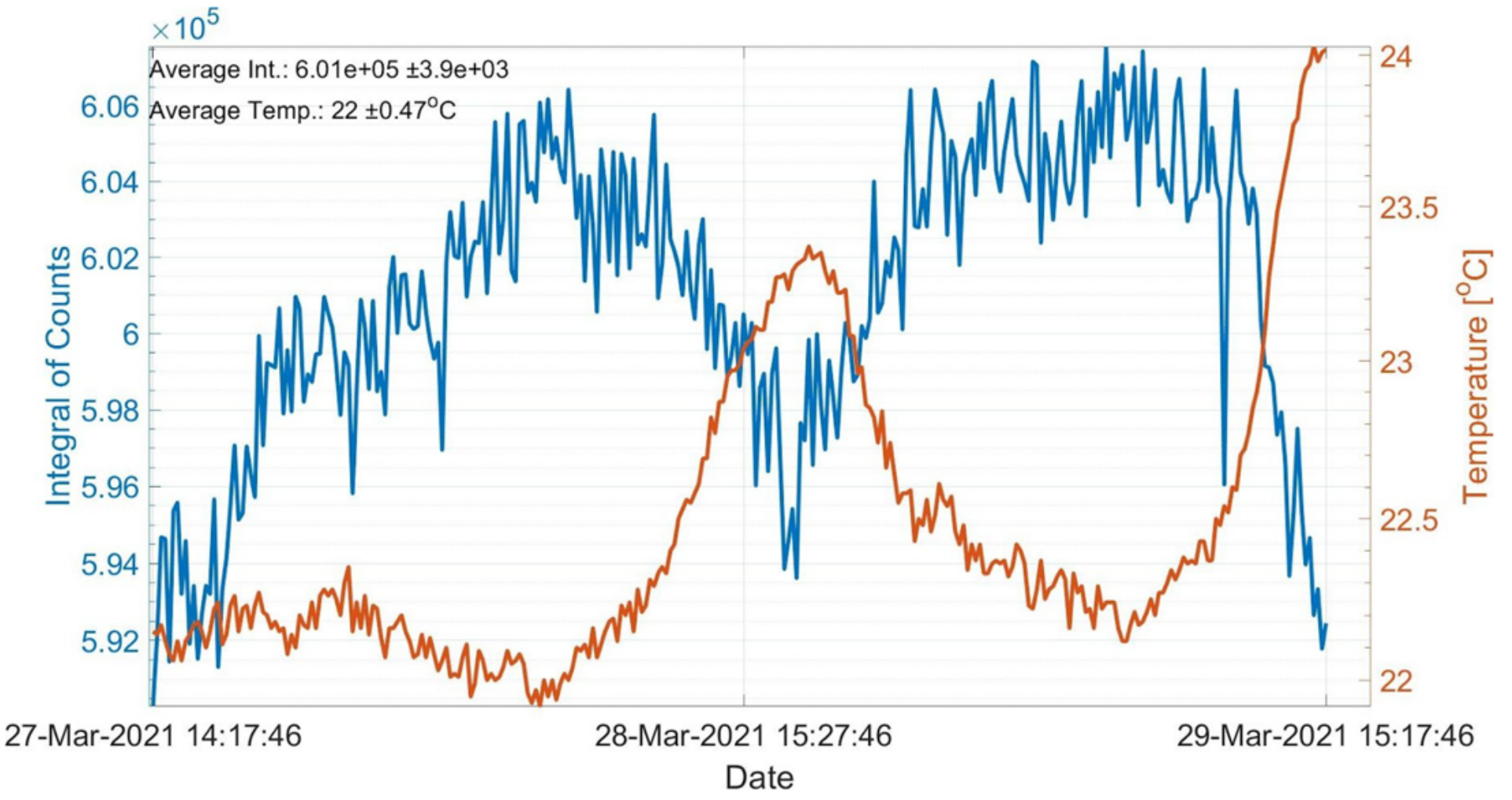

**Figure 4.** The variation of the integral of the intensity for the UV LED and the room temperature for two consecutive days.

The optical transmittance of the glass samples were measured in 200 nm – 1500 nm range with the Shimadzu UV-3600 Plus UV-VIS-NIR Spectrophotometer [7] for more than five months, starting with daily measurements in the first two months and with decreasing frequency afterwards. In addition, the optical transmittance of an unirradiated glass sample of the same type was acquired for the relative transmittance calculations. Successive measurements of transmittance yield less than 0.1 % variation between measurements. The automated transition between the photodetectors of the spectrophotometer resulted in a slight non-uniformity in the measured spectra at the transition wavelengths, and the effect had to be corrected offline. The associated uncertainty due to this correction is less than 4.2 %.

## 3. Results

Figure 5 shows all the spectra overlayed for 3.5 kGy (top row) and 7.0 kGy (bottom row) for dark box (column 1) and ambient light (column 2) conditions, and white LED (column 3), UV LED (column 4), blue LED (column 5), green LED (column 6) and red LED (column 7) stimulations. The spectra measured immediately following the irradiation is clearly visible in all sample points, and the recovery in the first few days after the irradiation is dramatically high. The UV LED stimulation manifests itself with fast and large recovery followed by the blue LED stimulation.

Figure 6 shows the spectral dynamics of recovery from radiation damage in 340 – 1000 nm range, where the majority and the most relevant part of the radiation damage and recovery occurs, for the irradiated samples in the same order as in Fig. 5. Each slice parallel to the horizontal axis shows the ratio of the transmittance spectrum of the corresponding time on the vertical axis to the transmittance spectrum of the unirradiated sample, hereby called the relative transmittance. For UV LED stimulated recovery, the relative transmittance improves beyond 80 % for the entire spectral range in the final days of the recovery. For the other recovery modes, the 80 % relative transmittance threshold lies between 500 nm and 700 nm. The initial relative transmittance following the irradiation beyond 700 nm is around 80 % and the improvement in this range is minimal for all recovery modes.

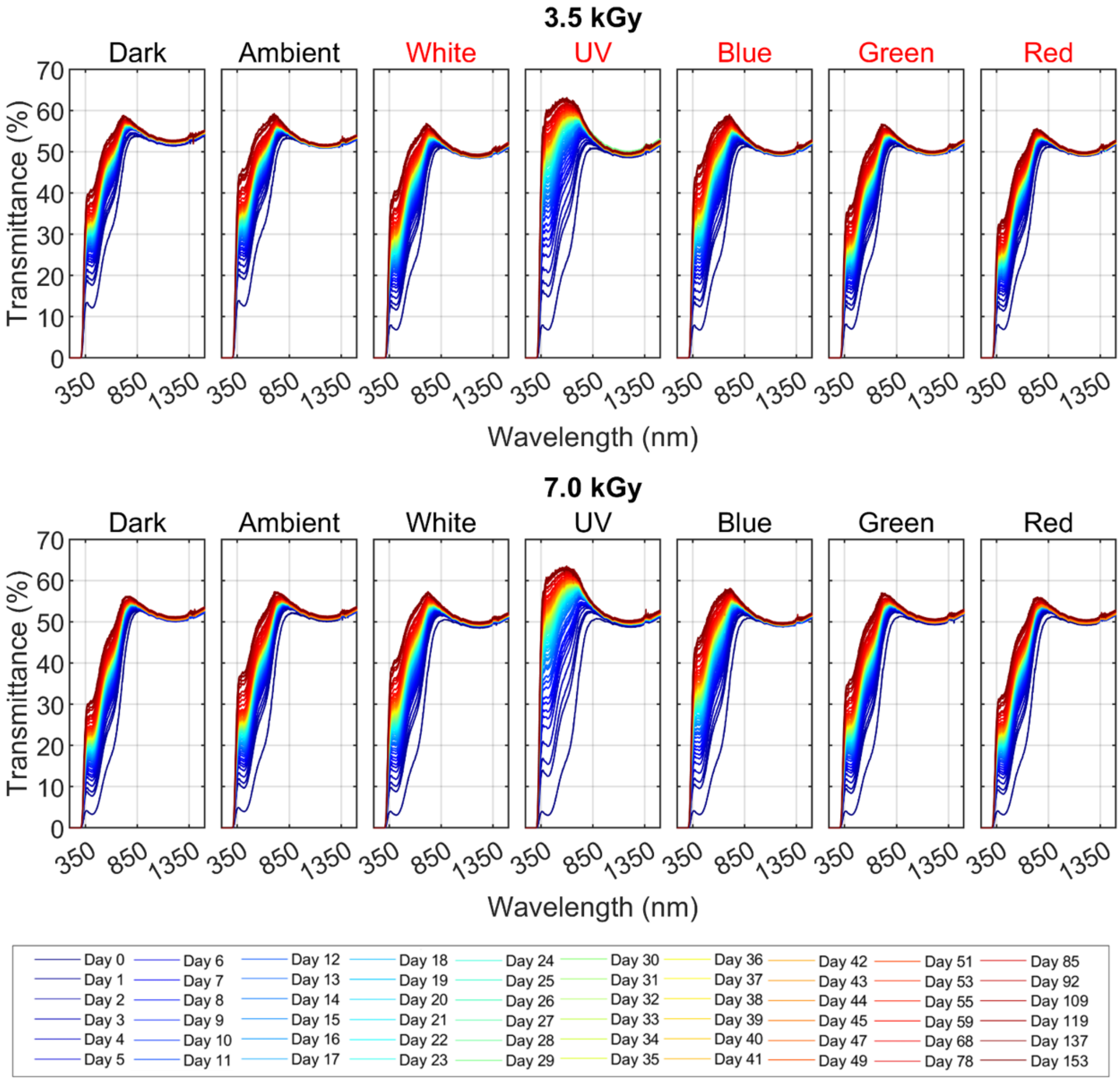

**Figure 5.** Overlayed spectra for 3.5 kGy (top row) and 7.0 kGy (bottom row) irradiations for dark box (column 1) and ambient light (column 2) conditions, and white LED (column 3), UV LED (column 4), blue LED (column 5), green LED (column 6) and red LED (column 7) stimulations.

The relative transmittance spectra were integrated in 340 – 1000 nm range in order to calculate the integrated transmittance loss (ITL). Figure 7 shows the ITL as a function of time after the irradiation for 3.5 kGy (left) and 7.0 kGy (right) measurement points. The LED stimulated recovery clearly follows an increasing performance in the order of red, green, white, blue and UV LEDs for both irradiation levels. For 7.0 kGy samples, the red LED stimulation is measured to be compatible with dark box condition, the green LED stimulation to be slightly worse than the ambient light condition and the white LED stimulation to be slightly better than the ambient light condition. The significant improvement beyond natural recovery is observed for stimulation with shorter wavelength LEDs, namely the blue and the UV LEDs. The difference between the thicknesses of the LED stimulated sample and the samples recovered in the dark and ambient light conditions for 3.5 kGy irradiation is manifest both in terms of initial damage and recovery regimes.

The data points were fit to the sum of two exponentials and a constant, $Ae^{-t/\tau_{\text{fast}}} + Be^{-t/\tau_{\text{slow}}} + C$; where $t$ is the time after irradiation, $\tau_{\text{fast}}$ is the fast component of the recovery to quantify the dramatic increase in the relative transmittance in the first few days following the irradiation, $\tau_{\text{slow}}$ is the slow component of recovery to quantify the long term effect of the various recovery mechanisms probed, $C$ is the permanent damage, and $A$ and $B$ are free scaling parameters. Table 2 shows the results of the fits.

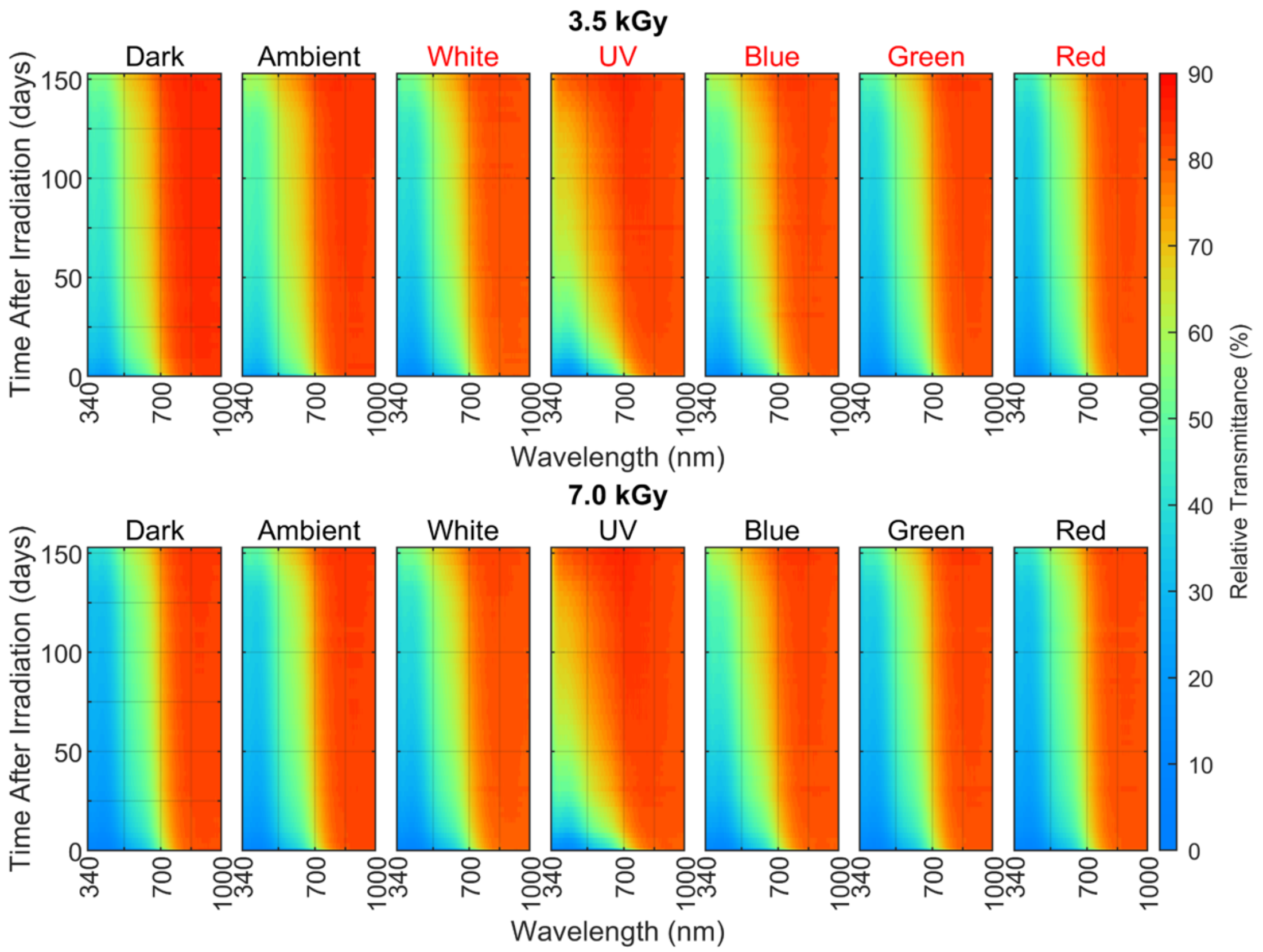

**Figure 6.** The spectral dynamics of recovery from radiation damage in 340 – 1000 nm range for 3.5 kGy (top row) and 7.0 kGy (bottom row) for dark box (column 1) and ambient light (column 2) conditions, and white LED (column 3), UV LED (column 4), blue LED (column 5), green LED (column 6) and red LED (column 7) stimulations.

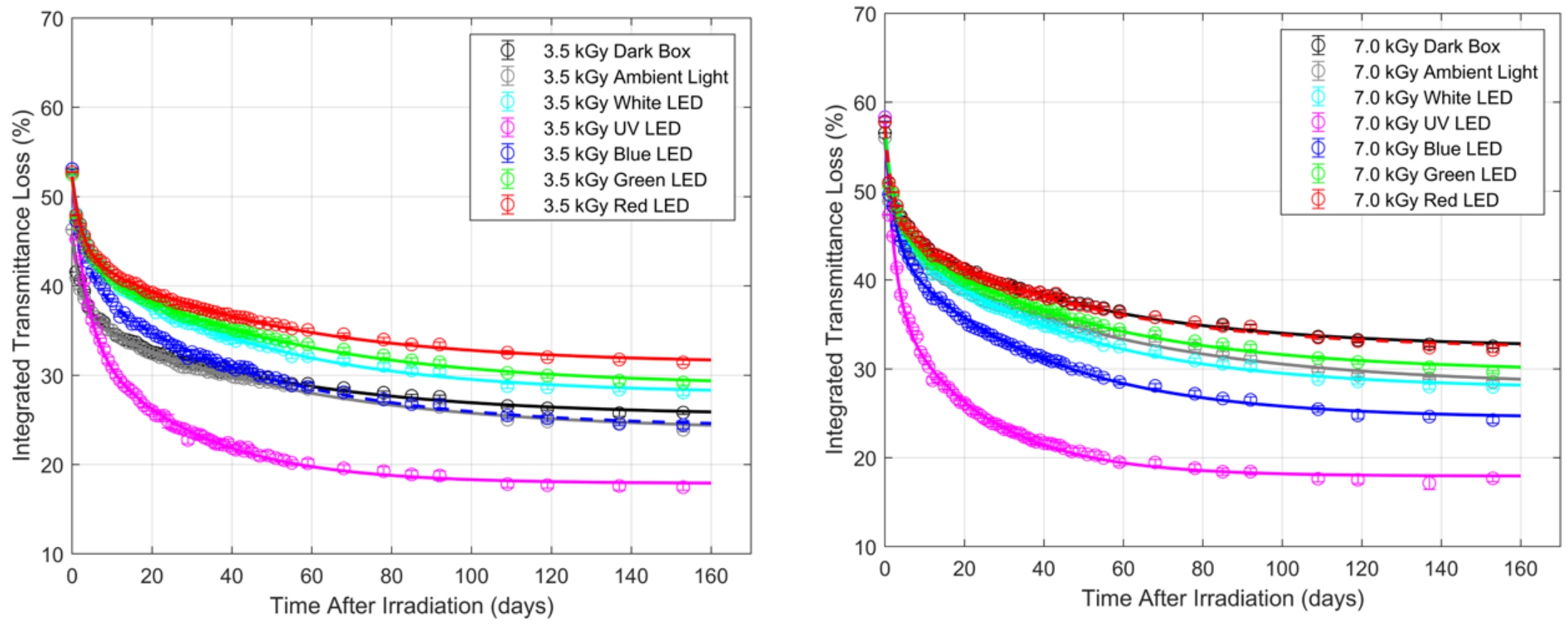

**Figure 7.** The integrated transmittance loss as a function of time after the irradiation for 3.5 kGy (left) and 7.0 kGy (right) measurement points. Color code for the curves and points are black for dark box and grey for ambient light conditions; and light blue for while LED, purple for UV LED, blue for blue LED, green for green LED and red for red LED stimulations. Curves are the fits to the points (see text for details). 3.5 kGy blue LED and 7.0 kGy red LED fit functions were drawn as dashed lines for better visibility of the underlying curves.

The fast component of recovery, $\tau_{\text{fast}}$, is within 1.5 days and 3.5 days for all recovery modes. It should be noted here that the LED stimulated recovery of 3.5 kGy irradiated sample should be considered separately due to the higher thickness of the sample. For both 3.5 kGy and 7.0 kGy irradiated samples, the UV and blue stimulated sample points show larger fast components compared to the other LED stimulated sample points. This is attributed to the superior long term recovery performance which gets closer to the fast recovery time frame and the associated difficulty in precisely distinguishing between the fast and slow recovery processes. This mixing of the time frames is most manifest at the comparable scaling factors A and B for the fast and slow recovery components for the UV LED stimulated samples.

**Table 2.** The parameters of the fits shown in Fig. 7.

| | Name | A (%) | $\tau_{\text{fast}}$ (days) | B (%) | $\tau_{\text{slow}}$ (days) | C (%) |
|---|---|---|---|---|---|---|
| 3.5 kGy | **Dark** | 9.31 ± 0.50 | 2.78 ± 0.31 | 11.07 ± 0.35 | 49.47 ± 4.95 | 25.45 ± 0.41 |
| | **Ambient Light** | 9.03 ± 0.78 | 2.26 ± 0.40 | 12.96 ± 0.56 | 53.65 ± 7.08 | 23.74 ± 0.68 |
| | **White LED** | 8.91 ± 0.71 | 2.74 ± 0.44 | 15.62 ± 0.47 | 44.29 ± 4.00 | 27.89 ± 0.48 |
| | **UV LED** | 16.91 ± 1.13 | 3.34 ± 0.40 | 17.35 ± 0.95 | 27.06 ± 2.56 | 17.88 ± 0.36 |
| | **Blue LED** | 11.47 ± 0.81 | 3.35 ± 0.48 | 16.54 ± 0.57 | 44.98 ± 4.71 | 24.13 ± 0.57 |
| | **Green LED** | 8.98 ± 0.54 | 2.76 ± 0.34 | 14.13 ± 0.38 | 50.65 ± 4.35 | 28.78 ± 0.45 |
| | **Red LED** | 8.60 ± 0.52 | 2.50 ± 0.31 | 12.35 ± 0.34 | 47.45 ± 4.08 | 31.29 ± 0.39 |
| 7.0 kGy | **Dark** | 9.78 ± 0.64 | 2.53 ± 0.34 | 13.75 ± 0.42 | 46.89 ± 4.44 | 32.38 ± 0.47 |
| | **Ambient Light** | 11.23 ± 0.89 | 1.95 ± 0.31 | 16.19 ± 0.64 | 53.90 ± 6.25 | 28.00 ± 0.77 |
| | **White LED** | 12.02 ± 0.79 | 2.17 ± 0.29 | 17.92 ± 0.49 | 43.24 ± 3.47 | 27.72 ± 0.51 |
| | **UV LED** | 19.77 ± 1.16 | 2.34 ± 0.26 | 19.87 ± 0.88 | 23.14 ± 1.83 | 17.93 ± 0.35 |
| | **Blue LED** | 13.46 ± 0.86 | 2.28 ± 0.30 | 19.21 ± 0.53 | 37.80 ± 2.87 | 24.43 ± 0.48 |
| | **Green LED** | 11.20 ± 0.82 | 2.21 ± 0.33 | 16.15 ± 0.53 | 47.24 ± 4.68 | 29.65 ± 0.60 |
| | **Red LED** | 10.85 ± 0.76 | 1.96 ± 0.28 | 14.20 ± 0.47 | 45.59 ± 4.49 | 32.24 ± 0.52 |

The slow component of recovery, $\tau_{\text{slow}}$, and the permanent damage, $C$, are greatly improved for the UV LED stimulated samples. For 7.0 kGy irradiated samples, $\tau_{\text{slow}}$ is shorter by around 50 % (23.14 days versus 46.89 days) and the permanent damage is less by 45 % (17.93 % versus 32.38 %) for the UV LED stimulated recovery compared to the dark box recovery. An improvement, although to a lesser extent, is also observed for the blue LED stimulated recovery compared to the dark box recovery. There is not a remarkable difference among the other LED stimulated samples. For the case of the 7.0 kGy sample points, white, green and red LED stimulated recovery is comparable with the recovery described by the dark box and ambient light conditions. For the 3.5 kGy sample, $\tau_{\text{slow}}$ is shorter (27.06 days versus 47.45 days) and the permanent damage is less (17.88 % versus 31.29 %) by 43 % for the UV LED stimulated recovery compared to the red LED stimulated recovery. Since the red LED stimulated recovery is expected to be comparable to the dark box recovery condition, it can be concluded that 50 % lower dose does not result in significantly improved recovery due to higher thickness, which will be investigated further below.

Figure 8 shows $\tau_{\text{fast}}$ (top), $\tau_{\text{slow}}$ (center) and the permanent damage (bottom) as a function of the stimulating wavelength for UV, blue, green and red LEDs. The overall tendency for the fast component of the recovery is towards higher values for shorter stimulating wavelengths. This is mostly due to the slow component of the recovery getting closer to the fast component as the wavelength decreases. The fast component of recovery does not show significant variation across the wavelengths probed with the exception of a slight step change between 460 nm and 520 nm for the 3.5 kGy irradiated sample.

Starting from the blue wavelength, $\tau_{\text{slow}}$ drops dramatically, finally decreasing by approximately 50 % at 396 nm stimulating wavelength. The trend is clear and points towards further improvements with stimulating wavelengths deeper in the UV range. The slight increase in the green wavelength is possibly due to the intrinsic enhanced transparency of the glass to green light.

The permanent damage has a clear trend as a function of the stimulating wavelength, decreasing with decreasing wavelength. Compared to the wavelength of the red LED, 632 nm, the permanent damage is improved by 43 - 44 % for the UV wavelength, 396 nm. The trend is such that wavelengths deeper in the UV range are expected to lower the permanent damage further.

In order to investigate the effect of the sample thickness on the LED stimulated recovery, the ratio of the 7.0 kGy recovery trends to the 3.5 kGy recovery trends is studied. Figure 9 shows the ratio of the ITLs for these two irradiation levels. The ratios were fit to constant lines beyond 4 days where the fast recovery effects become less pronounced. The dark box and ambient light samples have comparable thicknesses whereas the 3.5 kGy LED station sample is approximately 33 % thicker than the 7.0 kGy LED station sample. The ITL ratio for the dark box samples is 1.26 and the ratio for the ambient light samples is 1.20. Therefore, the expected ratio due to the difference in the irradiation dose is expected to be in the range of around 20 % and 26 %. The ratios for the white, UV, blue, green and red LED stimulated sample points are 1.03, 1.00, 1.02, 1.04 and 1.04 respectively. Therefore, the recovery for the thicker sample is 15 – 20 % lower than the expected recovery from a sample with identical thickness. Particularly for the UV LED stimulation, 50 % lower dose is equivalent to 33 % higher thickness in terms of the recovery characteristics. This result points towards the feasibility of utilizing thinner optical active media in higher dose environments, also from the point of recovery procedures. It should be noted for Fig. 9 that the ratios of the recovery trends are constant for the entire measurement period of 5 months with $\chi^2$/ndf varying between 0.1 and 0.6; the ones for the UV and blue stimulation ratios being on the higher end. It can be concluded that the recovery regimes can be projected for different irradiation scenarios and sample thicknesses as long as the material type is kept constant. The effect of the dose rate on the recovery dynamics should be investigated separately.

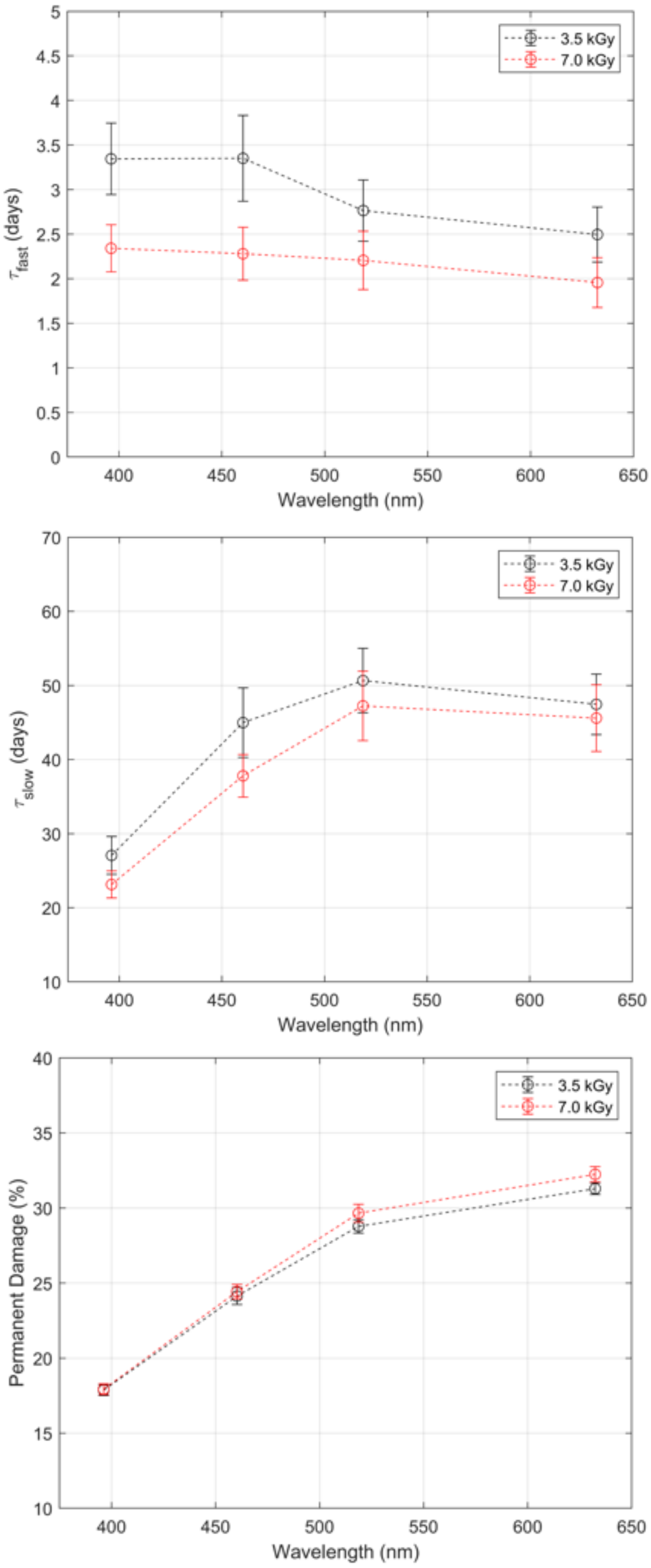

**Figure 8.** $\tau_{fast}$ (top), $\tau_{slow}$ (center) and the permanent damage (bottom) as a function of the stimulating wavelength for UV, blue, green and red LEDs.

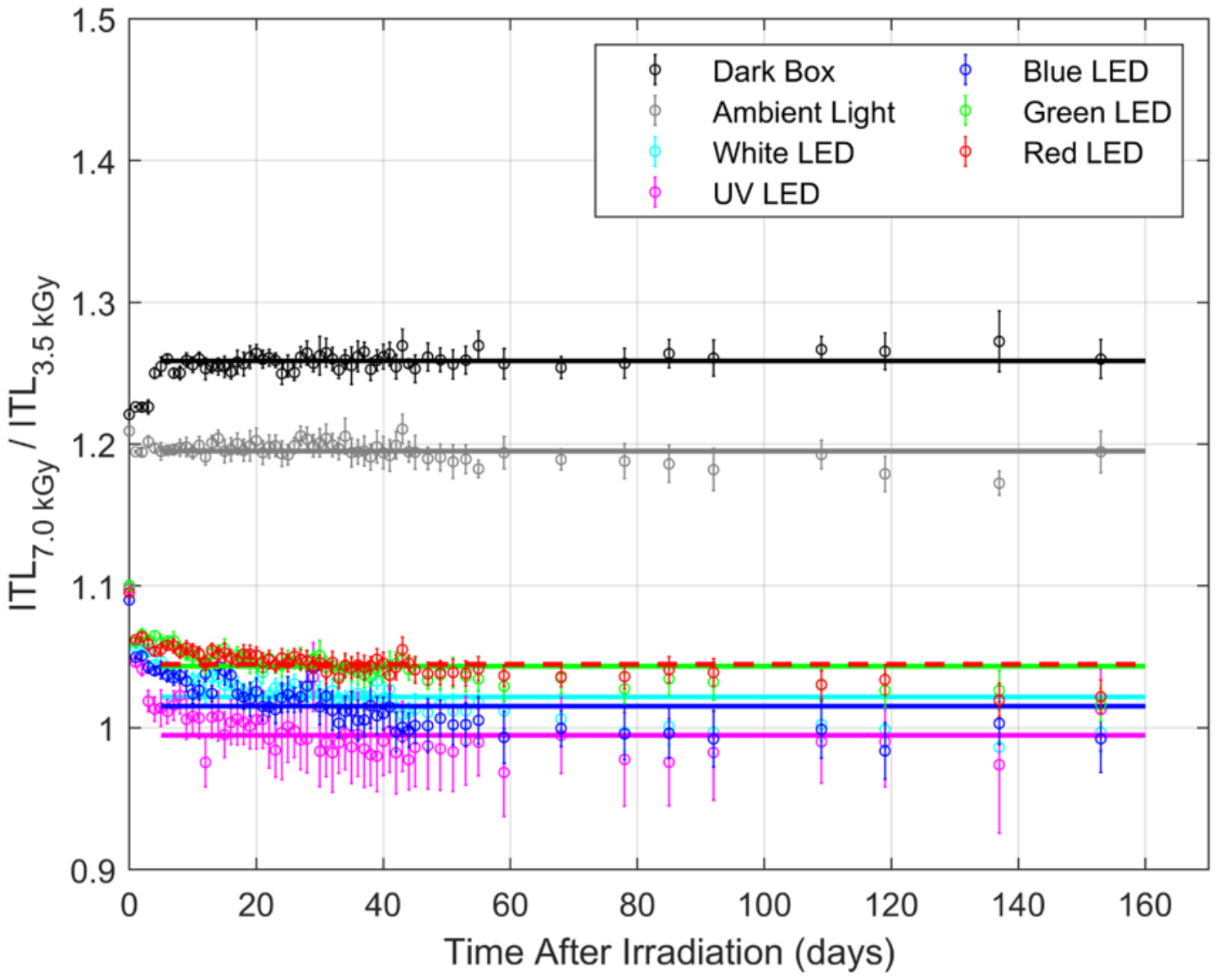

**Figure 9.** The ratios of the ITLs for the 7.0 kGy and 3.5 kGy samples and fits to a constant. The fit for the ITL ratio of the red LED stimulation is shown as a dashed line for better visibility of the underlying green line.

Figure 10 shows the fractional recovery spectra of the relative transmittance in different time domains and irradiation levels. Figure 10 top row shows the fractional recovery spectra in the first four days following the irradiation for 3.5 kGy (left) and 7.0 kGy (right) sample points. The spectral recovery has a characteristic shape with two major peaks, one around 360 nm and the other around 680 nm. The shorter wavelength peak extends from 340 nm to 420 nm. Apart from the observed ordering of UV, blue, white, green and red LED stimulation, the recovery shapes are in compliance with each other. As expected, the recovery shapes for the dark box and ambient light conditions exhibit a different regime compared to the LED stimulated recovery shapes for 3.5 kGy irradiation due to the difference in the sample thicknesses. On the other hand, the main features of this short-term recovery period are visible.

Figure 10 middle row shows the fractional recovery spectra from day 4 to day 40 for 3.5 kGy (left) and 7.0 kGy (right) sample points. The 360 nm peak is still visible. The UV stimulated recovery has an additional and more pronounced peak around 440 nm, which extends up to 540 nm. This peak is missing in all other fractional recovery spectra including the one with the blue LED stimulation. The characteristic ordering is still present but the distinction between different recovery mechanisms is more pronounced in this central time frame of 4 – 40 days. The fractional recoveries for the dark box and the red LED stimulation, and the ambient light condition and the green LED stimulation are comparable for this time frame for 7.0 kGy irradiation. Compared to the expectation based on this result, the thinner dark box and ambient light condition recovery samples exhibit an enhanced recovery in 340 nm – 600 nm range for the 3.5 kGy irradiation.

Figure 10 bottom row shows the fractional recovery spectra from day 40 to day 120 for 3.5 kGy (left) and 7.0 kGy (right) sample points. The 360 nm peak is not pronounced in any of the recovery regimes. The recovery in the ambient light condition dominates over the green LED stimulation in this time frame for 7.0 kGy irradiation. The 440 nm peak of the UV stimulated recovery is still visible, and is the dominant feature in the recovery curves of this time frame of 40 – 120 days. The 7.0 kGy fractional recovery curve of UV stimulated recovery shows a dramatic suppression for wavelengths larger than 540 nm. This is attributed to the saturation of the recovery. The saturation effect is also visible for the 3.5 kGy irradiated sample stimulated with UV LED. The recovery process is ongoing for the other recovery modes but less effectively and with the larger time constants shown in Table 2. The enhancement of the recovery in the 340 nm – 600 nm range due to lower thickness for the 3.5 kGy irradiation dark and ambient light condition samples is still manifest. In the longer time frame of recovery from radiation damage, the ambient light condition for a thinner optical sample becomes comparable with the blue LED stimulation for a thicker optical sample irradiated to the same dose.

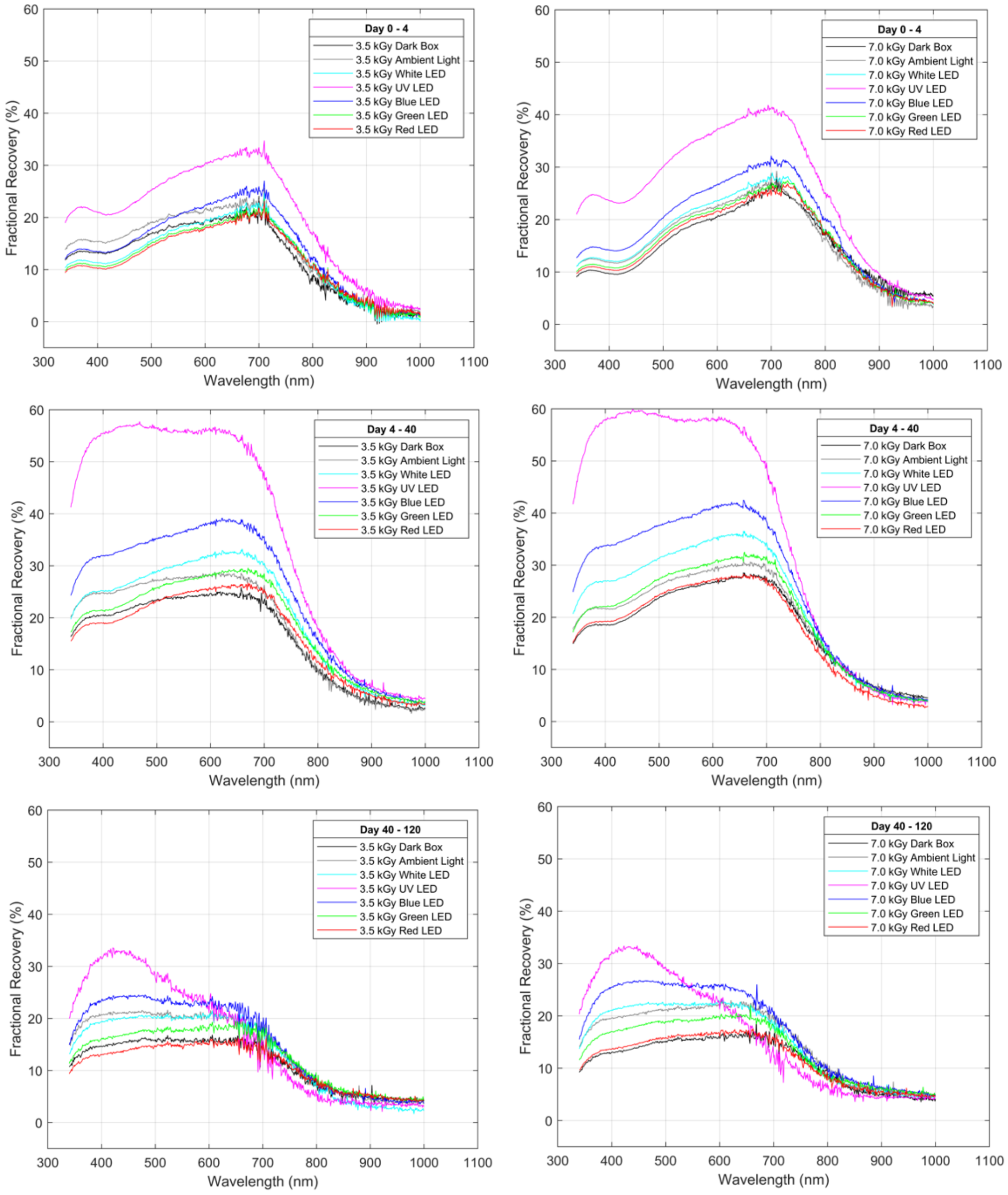

**Figure 10.** The fractional recovery spectra of the relative transmittance. Left column 3.5 kGy, right column 7.0 kGy; top row: day 0 to day 4; middle row: day 4 to day 40, bottom row: day 40 to day 120.

## 4. Systematic Uncertainties

The variation of the dose rate as a function of the position of the center line of the glass plates on the test table was calculated to obtain the systematic uncertainty on the total dose. For 7.0 kGy, the systematic uncertainty on the total dose was 4.2 %, and for 3.5 kGy, it was 3.9 %. These values were not used to make corrections of the total doses of the individual glass samples or in the calculation of the systematic errors of the results. But the dose rates quoted herein can be expressed as 7.0 ± 0.3 kGy and 3.5 ± 0.1 kGy.

The systematic uncertainties on the results were calculated for the possible sources of uncertainties and were added in quadrature to obtain a single value. The successive measurements of transmittance of the same sample to obtain a measure of the repeatability of the experiment and an upper bound of possible human error yielded an uncertainty of less than 0.1 %. The corrections implemented to account for the slight miscalibration of the photodetectors of the spectrophotometer resulted in a 4.2 % uncertainty on the results.

The overall systematic uncertainty was obtained as 4.2 % on individual measured spectra. As an upper limit of systematic uncertainties, this value can be added to the fit results of Table 2.

## 5. Summary

In order to perform systematic studies of recovery from radiation damage under various conditions, we irradiated three pieces of soda-lime float glass samples to total doses of 7.0 ± 0.3 kGy and 3.5 ± 0.1 kGy at a rate of 87.5 Gy/min. Immediately following the irradiation, one sample from each set was placed: in a dark box; in a room with ambient light; at the LED recovery station which provided pulsed LED stimulation at white, UV, blue, green and red wavelengths. The optical transmittance of the samples were measured in 200 – 1500 nm range for an extended period of time. The integral of the ratios of the transmittance spectra to the transmittance spectrum of the corresponding unirradiated sample were calculated in the range of 340 – 1000 nm, where the majority and the most relevant radiation damage and recovery occurs. The trend in the integrated transmittance loss was then investigated as a function of time after the irradiation. The trends were fit to the sum of two exponentials and a constant, $Ae^{-t/\tau_{\text{fast}}} + Be^{-t/\tau_{\text{slow}}} + C$. For 7.0 ± 0.3 kGy irradiation, the fast and slow time constants are 2.53 ± 0.34 (stat.) ± 0.11 (sys.) days and 46.89 ± 4.44 (stat.) ± 1.97 (sys.) days for the dark box recovery condition. The highest level of recovery was observed for UV LED stimulation for which the fast and slow time constants were 2.34 ± 0.26 (stat.) ± 0.10 (sys.) days and 23.14 ± 1.83 (stat.) ± 0.97 (sys.) days. The permanent damage for the dark box and UV LED stimulation recovery conditions was 32.38 ± 0.47 (stat.) ± 1.36 (sys.) % and 17.93 ± 0.35 (stat.) ± 0.75 (sys.) %. Therefore, UV LED stimulation results in 51 % better recovery rate with 45 % better total recovery compared to dark box recovery conditions. The blue LED stimulated recovery results in 19 % better recovery rate and 25 % total recovery compared to dark box recovery conditions. The red LED stimulation results are comparable with the dark box recovery results, and the green LED stimulation results are better than the dark box recovery results but still worse than the ambient light condition and the white LED stimulation results. These findings indicate that the stimulating wavelengths larger than 580 nm have no effect on the recovery from radiation damage and 500 – 580 nm range is minimally effective. The measurable difference in the recovery from radiation damage occurs at stimulating wavelengths shorter than 500 nm. The slow recovery time constant and the permanent damage show clear decreasing trends for decreasing wavelength of stimulating light. The results indicate that with wavelengths deeper in the UV range, better recovery conditions can be established.

The 3.5 ± 0.1 kGy irradiated LED stimulated sample was thicker than the others by 33 %. Therefore, the combined effect of different total dose and thickness on the recovery processes could be investigated simultaneously. The effect of the higher dose on the dark box and ambient light recovery was measured to be 26 % and 20 % respectively. The effect on the thicker sample on the other hand was measured to be between 0 % and 4 %. The difference of approximately 15 % was attributed to the higher thickness. Therefore, optical active media with lower thickness must be considered for high dose implementations, not only to obtain better intrinsic radiation-hardness but also to obtain more effective recovery from radiation damage. The near-perfect scaling of the recovery with the total dose and the sample thickness indicates that the performance of a range of implementations can be predicted with the data obtained as long as the sample material is the same. The extent of this range, the generalization of this principle to various other types of sample materials and the effect of the dose rate should be studied separately.

During fast recovery, i.e. within the first few days of recovery, all modes enhance recovery in 340 – 420 nm with a peak at 360 nm. In the medium time frame, from 4 to 40 days after irradiation, 340 – 540 nm range with a peak at 440 nm was observed to have higher recovery rate for the UV LED stimulation mode. This effect is not visible in the other LED stimulation modes, dark box and ambient light conditions. For the time frame of 40 to 120 days, the relatively enhanced recovery in 340 – 540 nm range is visible in the UV LED stimulation, and to a much lesser extent in the other LED stimulation modes. In this latest time frame, the suppression of the recovery by the UV LED stimulation for wavelengths larger than 540 nm was observed. This is attributed to the saturation of the recovery. Since the overall recovery by blue LED stimulation (and also the other modes of recovery to some extent) in this time frame still continues at a slower rate compared to the UV LED stimulation, the saturation effect is not pronounced as much. The spectral recovery fraction for the ambient light condition approaches the blue LED stimulation in the longer term recovery period when thinner optical media are used for the former recovery mode compared to the ones used for the latter.

## 6. Conclusions

A systematic study of LED stimulated recovery from radiation damage was performed. The following findings were obtained:

- LED stimulated recovery from radiation damage is a feasible and simple to implement technique for optical active media of radiation and particle detectors operating at high radiation environments.
- Shorter stimulating wavelengths result in faster recovery and lower permanent damage.
- There is a cut off stimulating wavelength above which the recovery is minimal to zero.
- The recovery characteristics of other irradiation scenarios, such as varying total dose and sample thickness, can be projected utilizing the current results.
- At longer time frames, the spectral recovery saturates starting from the green – red region.

Future studies should concentrate on specialty materials such as scintillators with emission peaks in the blue, green and orange range in order to obtain the complete picture of LED stimulated recovery from radiation damage.

## Acknowledgments

This work is supported by Tübitak grant no 118C224.